\documentclass{article}

\usepackage[final]{neurips_2021}

\usepackage[utf8]{inputenc} 
\usepackage[T1]{fontenc}    
\usepackage{hyperref}       
\usepackage{url}            
\usepackage{booktabs}       
\usepackage{amsfonts}       
\usepackage{nicefrac}       
\usepackage{microtype}      
\usepackage{xcolor}         

\usepackage{natbib}
\setcitestyle{authoryear,open={(},close={)}}

\usepackage{graphicx}
\usepackage{xcolor}

\graphicspath{{./images/}}
\usepackage{float}

\title{Molecule Generation from Input-Attributions
over Graph Convolutional Networks}

\author{%
 Dylan Savoia$^1$, Alessio Ragno\thanks{Correspondence to \href{mailto:ragno@diag.uniroma1.it}{ragno@diag.uniroma1.it}}\  $^1$, Roberto Capobianco$^{1,2}$\\\\
  $^1$Dept. of Computer, Control and Management \\Engineering ``Antonio Ruberti'', 
  Sapienza University of Rome \\$^2$Sony AI\\
}

\begin{document}

\maketitle

\begin{abstract}
It is well known that Drug Design is often a costly process both in terms of time and economic effort. While good Quantitative Structure-Activity Relationship models (QSAR) can help predicting molecular properties without the need to synthesize them, it is still required to come up with new molecules to be tested. This is mostly done in lack of tools to determine which modifications are more promising or which aspects of a molecule are more influential for the final activity/property. Here we present an automatic process which involves Graph Convolutional Network models and input-attribution methods to generate new molecules. We also explore the problems of over-optimization and applicability, recognizing them as two important aspects in the practical use of such automatic tools.  
\end{abstract}

\section{Introduction}

In this paper, we develop a novel molecular generation algorithm based on Graph Convolutional Networks (GCNs) that uses input-attribution scores, computed for each atom, to guide the generation process. GCNs are a type of neural network that accept graphs as their input and implement a convolutional operation similar to the one used in Convolutional Neural Networks: a learnable kernel is applied to each node and its neighbors in order to aggregate information among them and return meaningful features that can be used for classification or regression. Using input-attribution methods on graphs \textemdash{} which compute a score relative to how much each element of the input contributes to the outcome \textemdash{} we obtain a score for each atom that can then later be used by the generator.
The purpose of the generator is to compute a pool of molecules that are more likely to optimize some property of interest, in order to save resources in the first steps of drug-development and specifically in the context of searching for new candidates. The models required for the generation are developed on two datasets: first the Delaney water-solubility dataset used to develop the ESOL method \citep{delaney} which acts as a benchmark, and a second dataset of nearly 9000 molecules along with their activities against the \textit{Mycobacterium Tuberculosis} pathogen derived from the ChEMBL database \citep{chembl}. Both datasets express their molecules in the SMILES format \citep{smiles}: a string representation of their structure widely used in chemistry applications, used as an intermediate representation to exchange datasets of molecules.

The remainder of this paper will cover other work in the Chemistry literature that attempt to generate new molecules by different means; describe the proposed generator and finally go over the results of the generation process by taking into account different choices that can be made for the input-attribution scores and the type of additions that are tested during generation.

\section{Related Work}
First attempts to use graphs, are given by \citet{GraphKearnes} and \citet{3dgcn}. These develop architectures that also try to include 3D information as well. Notably the latter uses no absolute coordinates which is useful in keeping the representation invariant to geometric translations or rotations. More often, though, when considering graph representations we refer to a structure with no spatial information that only considers scalars, boolean or one-hot-encoding features.

The input-attribution scores required by the generator are computed through the methods known in the field of Explainable Artificial Intelligence (XAI). These methods can be as simple as computing the gradient of the output w.r.t. input and use the product between the gradient and the input as scores (InputXGradient), or rely on more complex techniques. \citet{IG} introduces ``Integrated Gradients'' (IG) which computes scores through the approximation of a path-integral while the ``DeepLIFT'' method proposed by \citet{deeplift} relies on a backpropagation-like algorithm. Both are solid methods that can be applied to a variety of different inputs and IG in particular also includes experiments performed on molecular tasks. These three methods \textemdash{} InputXGradient, IG and DeepLIFT \textemdash{} will be tested for the input-attribution component of the generator.

The generation of new molecules through machine learning methods can also take a variety of different approaches \citep{molgen_review}. One possibility is the direct generation of SMILES strings (used to represent the structure of molecules) through sequence models used in NLP (e.g. RNNs, LSTMs). One hybrid approach that uses reinforcement learning and LSTMs is the work done by \citet{release} and other projects use generative models such as Variational Auto-Encoders \citep{VAEgen} or Generative Adversarial Networks \citep{druGAN}.

\section{Proposed Generation Method}
We propose an algorithm capable of generating new molecules by making additions to already known ones. This mirrors the process followed by researchers when searching for new drug-candidates. Considering the process of finding a new drug for a pathogen, at first, molecules that are known to be active against a similar target are tested. Such molecules also act as a starting point for the generation of new ones, by searching for helpful additions that can improve their activity. 

The generator defined in this work requires a QSAR model in the form of a GCN; an input-attribution method and a set of possible additions (here referred to as `fragments') to be applied to an initial molecule. Given these elements the generator performs the following three steps: 
\begin{itemize}
    \item \textbf{Target Selection}: An atom belonging to the molecule is selected which will be the target of the modification. Specifically the atom with the lowest input-attribution score is selected in an attempt to improve the ``worst-performing'' part of the molecule.
    \item \textbf{Fragments Merging}: A set of generated molecules is created by applying a corresponding set of ``fragments'' and merging them with the initial molecule at the target atom location. 
    \item \textbf{Ranking by Model}: The resulting molecules are evaluated according to the QSAR model and ranked by their score. The first $N$ are selected to be shown to the end-user.
\end{itemize}

From this process we are able to generate new molecules given a starting one (a \textit{lead compound}) that is more likely to display activity or improved property when experimentally biologically testing the molecules (screening tests).

\paragraph{Fragments Choice and Generation:}
Any SMARTS fragment could be used: a notation similar to SMILES which includes the possibility to specify ``undefined'' atoms. Here two possibilities are explored: using a custom set of fragments made of functional groups originally recognized by \citet{FGroups} and fragments generated by the BRICS decomposition \citep{BRICS}. To exclude long fragments, these are filtered according to their length (in characters). For the Delaney dataset 35 fragments are obtained filtering out any SMARTS longer than 10 characters while for the \textit{M. Tuberculosis} dataset the limit is set to 20, obtaining 828 different fragments.

\section{Training and Results}
This section develops the building blocks of the generation method over two datasets: Delaney water-solubility (1128 molecuels) and the \textit{Mycobacterium Tuberculosis} dataset (8928 molecules) both in the SMILES format.
Water-solubility acts as a benchmark as simple evaluation rules from Chemistry are known. As an example, a chain of carbon atoms is hardly soluble in water because of its stability, but adding even a simple functional group can improve solubility (\citep{polar-solubility}).
Meanwhile, the property of interest of the \textit{Mycobacterium Tuberculosis} dataset is the activity against a pathogen, which makes it especially interesting in the context of drug-development.

\subsection{Training Setup}
We train a \textbf{GCN architecture} with different training hyper-parameters depending on the task, but all using as input a graph in which each node/atom is represented by: atom number, chirality, hybridization, atom mass, charge, radical electrons number, implicit valence and two boolean values that indicate whether the atom belongs to a ring and if the ring is aromatic. Minimization of the loss function is given by the \texttt{Adam} optimizer and the hardware on which models are trained includes an `RTX 2060 Super' GPU. During training the performance is evaluated on a Validation set and the best model overall is kept according to this evaluation.

For the Delaney dataset training occurred for 50 epochs and with a GPU memory footprint of around 1GiB to fully contain the model and the training and validation splits (of the 8GiB available). The initial learning rate for the \texttt{Adam} optimizer is set to 2e-3, while the size of the batch for each iteration is equal to 32.
For the \textit{Mycobacterium Tuberculosis} dataset, training occurred for 150 epochs using an bigger hidden representation of 512. We compare the GCN models with other learning algorithms used as a baseline, namely a Linear Regression model and a Random Forest Regressor as provided by the SK-Learn library. The models accept 1024-bit fingerprints of radius 2 computed through the RDKit Python library. Training for each class of model has been performed 50 times
on random train-test splits for all models and the results are summarized in Tables~\ref{tbl:SolMSE}.

\subsection {Model Results}
The results on the Delaney dataset show a significantly better performance of the GCN model compared to the alternatives, but this is not always the case: while a linear model is not sufficient in either of the two datasets, Random Forest performs well on the \textit{M. Tuberculosis} task. Despite the worse performance on the task, GCNs still present the important advantage of making input-attributions directly usable by being able to obtain scores for each atom. We will show later how this information can in fact improve the generation process.

\begin{table}[H]
    \centering
    \begin{tabular}{rcc}
        \textbf{Method} & \textbf{MSE Delaney} & \textbf{MSE M. Tuberculosis}\\
        Linear Regression:       & 1.542 $\pm$ 0.155 & 1.295 $\pm$ 0.045 \\
        Random Forest Regressor: & 1.441 $\pm$ 0.177 & 0.728 $\pm$ 0.036 \\
        Graph Conv. Networks:    & 0.671 $\pm$ 0.073 & 0.890 $\pm$ 0.038 \\
    \end{tabular}
    \caption{Training results for the Delaney and \textit{M. Tuberculosis} datasets obtained by different models.}
    \label{tbl:SolMSE}
\end{table}

\subsection{Generation Results}
\begin{figure}[t]
    \centering
    \includegraphics[width=\linewidth]{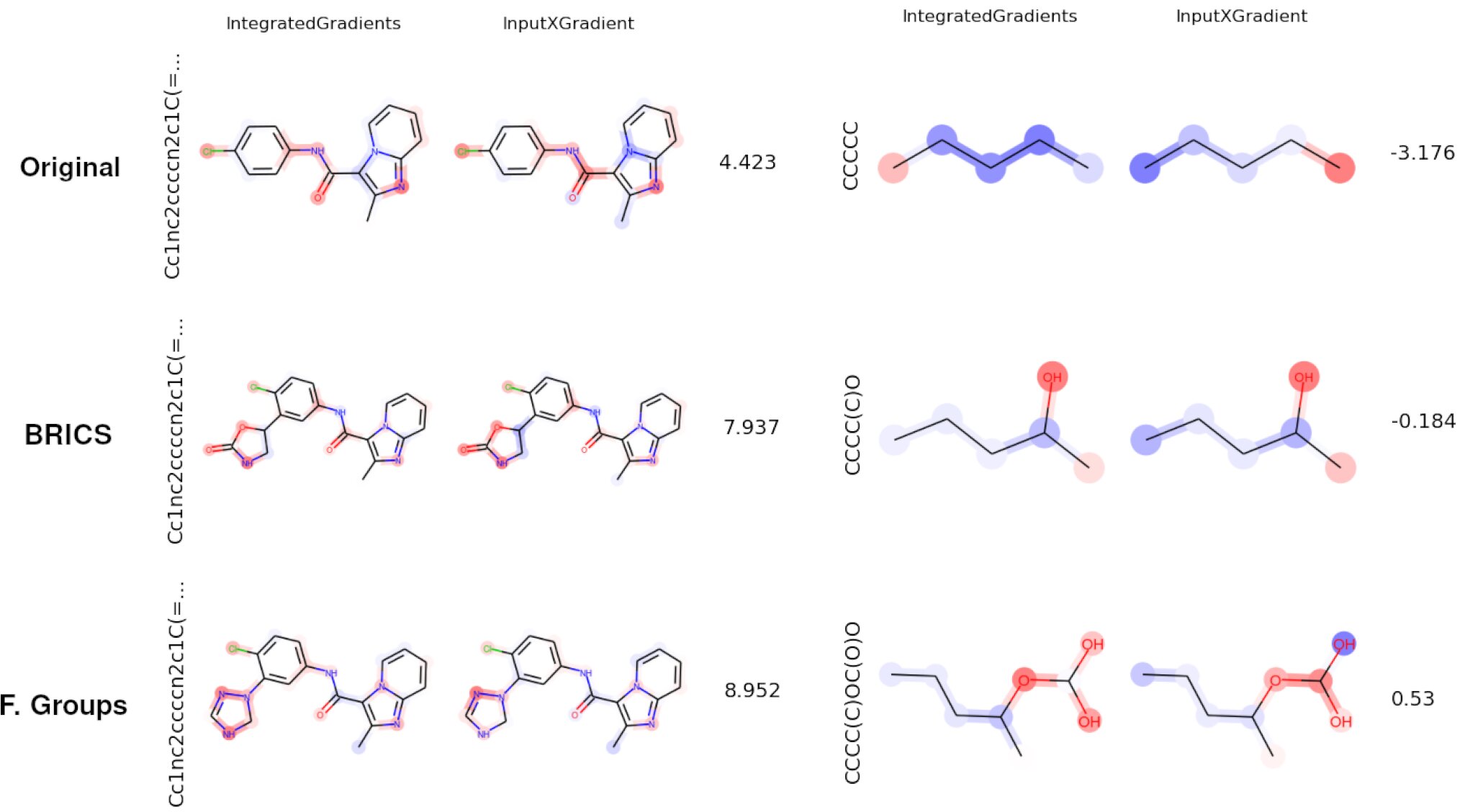}
    \caption{Generation example. Original molecule on top, then the best ones obtained using either BRICS or functional-groups fragments. (Left) generation starting from one molecule taken from the \textit{M. Tuberculosis} dataset (Right) generation of a highly soluble molecule starting from Pentane. For the input-attribution, \textbf{Red} represents a positive contribution to the output \textbf{Blue} a negative one.}
    \label{fig:genRes}
\end{figure}

An example of the results obtained by the Generator is provided in Figure~\ref{fig:genRes} in which  it is shown a tendency of functional groups of better exploring the optimization space. It can also be seen how the additions are correctly recognized by the input-attribution methods as positive contributions. This shows that the generator is in fact computing new molecules following the guidance of the GCN model.

\subsection{Repeated generation and Applicability}
Up until now the results obtained only considered one addition made by the generator. Because of this, we take 5 molecules from the \textit{M. Tuberculosis} dataset with  activity lower than 6 and repeat the generation process until any further addition stops obtaining improvements. Results show an activity increase up to 20/25, which is much higher than the highest value in the original dataset (12.0). Nevertheless we see that molecules have no chemical relevance: molecules are full of specific atoms (e.g. oxygen and nitrogen) suggesting an over-representation in the dataset of high-activity molecules containing these atoms or a fault in the GCN architecture.
Repeating the generation process also shows a limit of the GCN model to correctly assign an activity score to such molecules. We argue that the rising complexity of the molecules moves them away from the domain of applicability of the model making its predictions less useful. For these reasons, the recommended way of applying the generator presented in this work, is to start from a pool of randomly-selected known molecules and apply the generation on each one to obtain a new pool which is now more likely to contain active molecules. This contrasts with the possibility of generating new molecules from scratch, e.g. starting from a single carbon atom and build the rest of the molecule using the generator. Still, repeated generation may be used for debugging of the dataset distribution or model.

\section{Conclusions}
The current search of molecules as new drug candidates, involves a search of the molecular space which is mostly done by chance: big sets of molecules are synthesized with the hope that some of them can become drugs but not knowing much about their potential before they are tested. Our work proposes a way of generating molecules that takes into account the risks of the biases of automatic generation with the goal of developing a pool of candidate molecules that are more likely to display some property or activity of interest. In particular the process described here, assures that the results are chemically feasible and offers the possibility to generate molecules that do not differ too much from already known ones, making the predictions of the used QSAR models more trustworthy.

\medskip
\bibliography{bibliography} 

\begin{thebibliography}{14}
\providecommand{\natexlab}[1]{#1}
\providecommand{\url}[1]{\texttt{#1}}
\expandafter\ifx\csname urlstyle\endcsname\relax
  \providecommand{\doi}[1]{doi: #1}\else
  \providecommand{\doi}{doi: \begingroup \urlstyle{rm}\Url}\fi

\bibitem[Cho and Choi(2018)]{3dgcn}
H.~Cho and I.~S. Choi.
\newblock Three-dimensionally embedded graph convolutional network {(3DGCN)}
  for molecule interpretation.
\newblock \emph{CoRR}, abs/1811.09794, 2018.
\newblock URL \url{http://arxiv.org/abs/1811.09794}.

\bibitem[Degen et~al.(2008)Degen, Wegscheid-Gerlach, Zaliani, and Rarey]{BRICS}
J.~Degen, C.~Wegscheid-Gerlach, A.~Zaliani, and M.~Rarey.
\newblock On the art of compiling and using 'drug-like' chemical fragment
  spaces.
\newblock \emph{ChemMedChem}, 3:\penalty0 1503--7, 10 2008.
\newblock \doi{10.1002/cmdc.200800178}.

\bibitem[Delaney(2004)]{delaney}
J.~S. Delaney.
\newblock Esol: estimating aqueous solubility directly from molecular
  structure.
\newblock \emph{Journal of chemical information and computer sciences},
  44\penalty0 (3):\penalty0 1000--1005, 2004.

\bibitem[Ertl(2017)]{FGroups}
P.~Ertl.
\newblock An algorithm to identify functional groups in organic molecules.
\newblock \emph{Journal of Cheminformatics}, 9, 06 2017.
\newblock \doi{10.1186/s13321-017-0225-z}.

\bibitem[Gaulton et~al.(2012)Gaulton, Bellis, Bento, Chambers, Davies, Hersey,
  Light, McGlinchey, Michalovich, Al-Lazikani, et~al.]{chembl}
A.~Gaulton, L.~J. Bellis, A.~P. Bento, J.~Chambers, M.~Davies, A.~Hersey,
  Y.~Light, S.~McGlinchey, D.~Michalovich, B.~Al-Lazikani, et~al.
\newblock Chembl: a large-scale bioactivity database for drug discovery.
\newblock \emph{Nucleic acids research}, 40\penalty0 (D1):\penalty0
  D1100--D1107, 2012.

\bibitem[Kadurin et~al.(2017)Kadurin, Nikolenko, Khrabrov, Aliper, and
  Zhavoronkov]{druGAN}
A.~Kadurin, S.~Nikolenko, K.~Khrabrov, A.~Aliper, and A.~Zhavoronkov.
\newblock drugan: an advanced generative adversarial autoencoder model for de
  novo generation of new molecules with desired molecular properties in silico.
\newblock \emph{Molecular pharmaceutics}, 14\penalty0 (9):\penalty0 3098--3104,
  2017.

\bibitem[Kearnes et~al.(2016)Kearnes, McCloskey, Berndl, Pande, and
  Riley]{GraphKearnes}
S.~Kearnes, K.~McCloskey, M.~Berndl, V.~Pande, and P.~Riley.
\newblock Molecular graph convolutions: moving beyond fingerprints.
\newblock \emph{Journal of Computer-Aided Molecular Design}, 30\penalty0
  (8):\penalty0 595–608, Aug 2016.
\newblock ISSN 1573-4951.
\newblock \doi{10.1007/s10822-016-9938-8}.
\newblock URL \url{http://dx.doi.org/10.1007/s10822-016-9938-8}.

\bibitem[LibreTexts(2020)]{polar-solubility}
LibreTexts.
\newblock Solubility and factors affecting solubility.
\newblock 2020.
\newblock URL \url{https://chem.libretexts.org/@go/page/1613}.

\bibitem[Popova et~al.(2017)Popova, Isayev, and Tropsha]{release}
M.~Popova, O.~Isayev, and A.~Tropsha.
\newblock Deep reinforcement learning for de-novo drug design.
\newblock \emph{CoRR}, abs/1711.10907, 2017.
\newblock URL \url{http://arxiv.org/abs/1711.10907}.

\bibitem[Rigoni et~al.(2020)Rigoni, Navarin, and Sperduti]{VAEgen}
D.~Rigoni, N.~Navarin, and A.~Sperduti.
\newblock A systematic assessment of deep learning models for molecule
  generation.
\newblock \emph{CoRR}, abs/2008.09168, 2020.
\newblock URL \url{https://arxiv.org/abs/2008.09168}.

\bibitem[Shrikumar et~al.(2017)Shrikumar, Greenside, and Kundaje]{deeplift}
A.~Shrikumar, P.~Greenside, and A.~Kundaje.
\newblock Learning important features through propagating activation
  differences.
\newblock \emph{CoRR}, abs/1704.02685, 2017.
\newblock URL \url{http://arxiv.org/abs/1704.02685}.

\bibitem[Sundararajan et~al.(2017)Sundararajan, Taly, and Yan]{IG}
M.~Sundararajan, A.~Taly, and Q.~Yan.
\newblock Axiomatic attribution for deep networks.
\newblock \emph{CoRR}, abs/1703.01365, 2017.
\newblock URL \url{http://arxiv.org/abs/1703.01365}.

\bibitem[Weininger(1988)]{smiles}
D.~Weininger.
\newblock Smiles, a chemical language and information system. 1. introduction
  to methodology and encoding rules.
\newblock \emph{Journal of chemical information and computer sciences},
  28\penalty0 (1):\penalty0 31--36, 1988.

\bibitem[Xue et~al.(2019)Xue, Gong, Yang, Chuai, Qu, Shen, Yu, and
  Liu]{molgen_review}
D.~Xue, Y.~Gong, Z.~Yang, G.~Chuai, S.~Qu, A.~Shen, J.~Yu, and Q.~Liu.
\newblock Advances and challenges in deep generative models for de novo
  molecule generation.
\newblock \emph{Wiley Interdisciplinary Reviews: Computational Molecular
  Science}, 9\penalty0 (3):\penalty0 e1395, 2019.

\end{thebibliography}

\end{document}